\documentstyle[12pt]{article}

\begin{document}
{\it University of Shizuoka}

\hspace*{10.5cm} {\bf US-00-08}\\[-.3in]

\hspace*{10.5cm} {\bf August 2000}\\[-.3in]

\hspace*{10.5cm} {\bf hep-ph/0008129}\\[.3in]

\vspace*{.4in}

\begin{center}

{\Large\bf  Bimaximal Neutrino Mixing in a Zee-type Model\\[.2in]

with Badly Broken Flavor Symmetry} \\[.3in]

{\bf Yoshio Koide}\footnote{
E-mail: koide@u-shizuoka-ken.ac.jp} {\bf and
Ambar Ghosal}\footnote{
E-mail: gp1195@mail.a.u-shizuoka-ken.ac.jp} \\

Department of Physics, University of Shizuoka \\ 
52-1 Yada, Shizuoka 422-8526, Japan \\[.1in]

\vspace{.3in}

{\large\bf Abstract}\\[.1in]

\end{center}

\begin{quotation}
A Zee-type neutrino mass matrix model with a badly broken 
horizontal symmetry SU(3)$_H$ is investigated.
By putting a simple ansatz on the symmetry breaking 
effects of SU(3)$_H$ for transition matrix elements, 
it is demonstrated that the model can give
a nearly bimaximal neutrino mixing with the ratio
$\Delta m^2_{solar}/\Delta m^2_{atm}
\simeq \sqrt{2} m_e/m_{\mu}=6.7 \times 10^{-3}$,
which are in excellent agreement with the observed data.
In the near future, the lepton-number violating decay
$Z\rightarrow \mu^\pm \tau^\mp$ will be observed.
\end{quotation}

\vfill
\newpage
\section{Introduction}
The recent Super-Kamiokande collaboration \cite{solar-SK} has reported, 
by comparing the day/night spectrum and results of flux global analysis, 
that the small mixing angle MSW and just-so solutions for active neutrinos 
are disfavored at 95$\%$ C.L. and a mixing with sterile neutrinos is also 
disfavored at 95$\%$ C.L.  On the other hands, we have already known 
that the 
atmospheric neutrino data suggests a $\nu_{\mu}\leftrightarrow\nu_{\tau}$ 
mixing with a large mixing angle $\sin^{2}2\theta\simeq1$ \cite{atm}.  
If we take these experimental results seriously, we are forced 
to accept only a model which 
gives a nearly bimaximal mixing among the active neutrinos 
$(\nu_e, \nu_{\mu}, \nu_{\tau})$.  
We must seek for the origin of the nearly bimaximal mixing.

As promising one of such the models, the Zee model \cite{Zee} is known.  
In this model, a charged scalar field $h^+$ is introduced in addition 
to the Higgs doublets $\Phi_1$ and $\Phi_2$; 

$$
{\cal L}={\frac{1}{2}}\sum_{i,j}f_{ij}\overline{\ell}_{iL}i\tau_2\ell_{iL}^c h^-+
c_{12}\Phi^T_{1}i\tau_2\Phi_2h^-
+\sum_{i}y_i\overline{\ell}_{iL}\Phi_2 e_{iR}+h.c.
\eqno(1.1)
$$
$$
\ell_{iL}=
\left(\begin{array}{c}
\nu_{iL} \\
e_{iL}
\end{array}\right), \ \ 
\Phi_a=
\left(\begin{array}{c}
\Phi_a^+ \\
\Phi_a^0
\end{array}\right),\ \ 
i\tau_2=
\left(\begin{array}{cc}
0 & 1 \\
-1 & 0 
\end{array}\right),
\eqno(1.2)
$$
where $i=1,2,3$ are family indexes, 
$\ell^c_{iL}=(\ell_{iL})^c=C \overline{\ell}_{iL}^T$, and
$c_{12}=-c_{21}$ are real mass parameter.  The neutrino mass 
matrix $M$ is radiatively generated as 
$$
M_{ij}=m_0f_{ij}(m_{ej}^{2}-m_{ei}^{2})/m_{\tau}^2,
\eqno(1.3)
$$
where
$$
m_0=
\frac{\sin 2\phi \tan\beta m_\tau^2}{32\pi^2 v/\sqrt{2}}
\ln {\frac{M_1^2}{M_2^2}},
\eqno(1.4)
$$
$$
\sin 2\phi = - \frac{2 c_{12} v/\sqrt{2}}{M_1^2 -M_2^2} \ ,
\eqno(1.5)
$$
$m_{ei}=y_i \langle\Phi_2^0\rangle=
y_i(v/{\sqrt{2}})\cos\beta$, $\tan\beta=\langle\Phi^0_1
\rangle/\langle\Phi_2^0\rangle$, 
$v/\sqrt{2}=\sqrt{|\langle\Phi_1^0\rangle|^2+|\langle\Phi_2^0\rangle|^2}
= 174$ GeV, and 
$M_1$ and $M_2$ are the masses of the charged scalars $H_1^+$ and $H_2^+$, 
respectively, 
which are mass eigenstates of $(h^+, \Phi^+)$, i.e., 

$$
\left(\begin{array}{c}
H_1^+ \\
H_2^+
\end{array}\right)=
\left(\begin{array}{cc}
\cos\phi & -\sin\phi \\
\sin\phi & \cos\phi
\end{array}\right)
\left(\begin{array}{c}
h^+ \\
\Phi^+
\end{array}\right),
\eqno(1.6)
$$
$$
\Phi^+=\Phi_1^+\cos\beta-\Phi_2^+\sin\beta.
\eqno(1.7)
$$

Based on the recent solar and atmospheric neutrino data, 
Smirnov and Tanimoto \cite{Smirnov-Tanimoto} have investigated 
the model in detail, and they 
have concluded that there is no solution of the solar neutrino problem unless 
introducing a sterile neutrino $\nu_s$, although the model can explain the 
observed large mixing $\sin^{2}2\theta_{atm}\simeq1$.  However, they have 
investigated only the case with 
$\varepsilon=M_{12}/{\sqrt{M_{23}^2+M_{13}^2}}\ll1$ and 
$\tan\theta=M_{13}/M_{23}\ll1$ considering 
${\Delta}m_{21}^2\ll{\Delta}m_{32}^2\ (m_1<m_2<m_3)$, and they have not 
considered the case $|M_{12}|\sim|M_{13}|\gg|M_{23}|$.  
The case $|M_{12}|\sim|M_{13}|\gg|M_{23}|$ has been investigated by 
Jarlskog {\it et al}. \cite{Jarlskog}, and they have pointed out that the 
case can lead to the nearly bimaximal mixing.

In the present paper, we put a simple ansatz on the coupling constants 
$f_{ij}$ and $y_i$ under a badly broken flavor symmetry, and thereby we 
will obtain the nearly bimaximal mixing together with a prediction 
${\Delta}m^2_{solar}/{\Delta}m^2_{atm}\simeq{\sqrt{2}}m_e/m_{\mu}=
6.7\times10^{-3}$.

\section{Assumption on the symmetry breaking of SU(3)$_H$}

We consider a badly broken horizontal symmetry \cite{h-sym} SU(3)$_H$.  
We introduce parameters $s_i\ (i=1,2,3)$ as a measure of 
the symmetry breaking of SU(3)$_H$.
In the present paper, we do not touch the origin of the symmetry breaking.

Our basic assumption on the magnitudes of the symmetry breaking effects 
is as follows: 
The magnitude of the matrix element 
${\langle}e_i({\bf p})|y_{ij}(\overline{e}_ie_j)|e_j({\bf p})\rangle$, 
which is a component of $3\times3^{\ast}=1+8$ of SU(3)$_H$, is 
proportional to $\delta_{ij}s_{i}^2$ in the limit of 
$|{\bf p}|\rightarrow\infty$, 
i.e., 
$$
y_i{\langle}e_i({\bf p})|(\overline{e}_ie_i)|e_i({\bf p})\rangle=
s_i^2\times{const}, 
\eqno(2.1)
$$
while the magnitude of the matrix element $\langle\overline{\nu}_{iL}({\bf p})
|f_{ij}(\overline{\nu}_{iL}^ce_{jL})|e_j^-({\bf p})\rangle
-(i\leftrightarrow j)$, which is a component of $3^{\ast}$ of SU(3)$_H$, 
is proportional to $\Sigma_k\varepsilon_{ijk}s_k$ in the limit of 
$|{\bf p}|\rightarrow\infty$,  i.e.,
$$
f_{ij}\langle\overline{\nu}_{iL}({\bf p})|(\overline{\nu}_{iL}^ce_{jL})|
e_j^-({\bf p})\rangle 
-(i \leftrightarrow j) =
\sum_k\varepsilon_{ij}s_k\times{const}.
\eqno(2.2)
$$
Here, note that our requirements are applied in the limit of 
$|{\bf p}|\rightarrow\infty$, because the state $e_{iL}$ (
also even $\nu_{iL}$) is not eigenstate
of the helicity $h$ in finite momentum frame unless 
the particle is massless.

These matrix elements, the left-hand sides of Eqs.~(2.1) and (2.2),
are evaluated as follows:
$$
y_i {\frac{1}{(2\pi)^3}}
{\frac{m_{ei}}{E_{ei}}}
\overline{u}_{ei}({\bf p})u_{ei}({\bf p}),
\eqno(2.3)
$$
$$
f_{ij}{\frac{1}{(2\pi)^3}}
{\sqrt{\frac{m_{ej}}{E_{ej}}}}
{\sqrt{\frac{m_{\nu i}}{E_{\nu i}}}}
\overline{u}_{\nu iL}^{c}({\bf p}) u_{e jL}({\bf p}) 
-(i \leftrightarrow j) \ ,
\eqno(2.4)
$$
respectively, 
where the spinor $u_i({\bf p})$ is normalized as 
$\overline{u}_i({\bf p})u_i({\bf p})=1$.  
Since in the limit of $|{\bf p}|\rightarrow \infty$, 
we obtain $\overline{u}_{ei}({\bf p})u_{ei}({\bf p})=1$ and 
$$
\lim_{|{\bf p}|\rightarrow \infty} 
\overline{u}_{\nu iL}^{c}({\bf p})u_{e jL}({\bf p})=
\frac{m_{ej}}{2 \sqrt{m_{\nu i}m_{e j}} } \ ,
\eqno(2.5)
$$
so that the assumptions (2.1) and (2.2) require
$$
y_im_{e i}=s_i^2\times{const},
\eqno(2.6)
$$
and 
$$
f_{ij}(m_{ej}+m_{ei})=\sum_k\varepsilon_{ijk}s_k\times{const},
\eqno(2.7)
$$
respectively, where ``$const$" includes $|{\bf p}|$.

Note that even if we apply our ansatz to the terms
$\sum_i y_i (\overline{\nu}_{iL} e_{iR}) \Phi^+ +h.c.$
instead of the terms 
$\sum_i y_i (\overline{e}_{iL} e_{iR}) \Phi^0 +h.c.$ 
which leads to the requirement (2.1),
we can obtain the same result with the result (2.6)
because of
$$
y_i {\frac{1}{(2\pi)^3}}
\sqrt{\frac{m_{\nu i}}{E_{\nu i}}}\sqrt{\frac{m_{ei}}{E_{ei}}}
\overline{u}_{\nu i}({\bf p})u_{ei}({\bf p}),
\eqno(2.8)
$$
and 
$$
\lim_{|{\bf p}|\rightarrow \infty} 
\overline{u}_{\nu iL}({\bf p})u_{e iR}({\bf p})=
\frac{m_{ei}}{2 \sqrt{m_{\nu i}m_{e i}} } \ ,
\eqno(2.9)
$$

Recalling $y_i=m_{ei}/\langle\Phi_2^0\rangle$, 
from the results (2.6) and (2.7), 
we obtain the symmetry breaking effect on 
the coupling constants $f_{ij}$
$$
f_{ij}\propto
{\frac{\varepsilon_{ijk}s_k}{m_{ej} +m_{ei}}}\propto
{\frac{\varepsilon_{ijk}m_{ek}}{m_{ej} +m_{ei}}},
\eqno(2.10)
$$
i.e.,
$$
f_{e\mu}=\varepsilon_{123}
\frac{m_{\tau}}{m_{\mu} +m_e} f\ , \ \ 
f_{e\tau}=\varepsilon_{132}
 \frac{m_{\mu}}{m_{\tau} +m_e} f \ , \ \ 
f_{\mu\tau}=\varepsilon_{231}
\frac{m_e}{m_{\tau} +m_{\mu}} f \ .
\eqno(2.11)
$$

\section{Mass spectrum and bimaximal mixing}

By using the results (2.11), 
we obtain the following mass matrix elements $M_{ij}$:
$$
M_{e\mu}=\varepsilon_{123}m_{\tau}(m_{\mu}-m_e) f m_0/m_{\tau}^2 \ ,
$$
$$
M_{e\tau}=\varepsilon_{132}m_{\mu}(m_{\tau} -m_e) f m_0/m_{\tau}^2 \ ,
$$
$$
M_{\mu\tau}=\varepsilon_{231}m_{e}(m_{\tau}-m_{\mu}) f m_0/m_{\tau}^2 \ .
\eqno(3.1)
$$

Since $m_{\tau}{\gg}m_{\mu}{\gg}m_e$, we obtain
$$
M \simeq
\left(\begin{array}{ccc}
0 & a & -a \\
a & 0 & b \\
-a & b & 0 
\end{array}\right),
\eqno(3.2)
$$
where 
$$
a=
{\frac{m_{\mu}}{m_{\tau}}} f m_0 \ ,\ \ \ 
b=
{\frac{m_e}{m_{\tau}}} f m_0 \ .
\eqno(3.3)
$$
The matrix form (3.2), except for the sign of $M_{13}$, 
is identical with the neutrino mass matrix which has recently 
been proposed by one of the authors (A.G.) \cite{Ghosal} on the basis of 
discrete $Z_3{\times}Z_4$ symmetries, and it is know that the matrix form 
(3.2) can lead to the nearly bimaximal mixing.  
The mixing matrix $U$ and mass eigenvalues $m_{\nu i}$ are as follows: 
$$
U=
\left(\begin{array}{ccc}
\cos\theta & -\sin\theta & 0 \\
-{\frac{1}{\sqrt{2}}}\sin\theta & -{\frac{1}{\sqrt{2}}}\cos\theta & 
{\frac{1}{\sqrt{2}}} \\
{\frac{1}{\sqrt{2}}}\sin\theta & {\frac{1}{\sqrt{2}}}\cos\theta 
& {\frac{1}{\sqrt{2}}} 
\end{array}\right),
\eqno(3.4)
$$
$$
\tan\theta=
\left(
{\frac
{{\sqrt{8a^2+b^2}}+b}
{{\sqrt{8a^2+b^2}}-b}
} \right)^{1/2}
=
{\sqrt
{\frac{-m_{\nu 1}}
{m_{\nu 2}}
}},
\eqno(3.5)
$$
$$
m_{\nu 1}=-{\frac{1}{2}}
\left(
{\sqrt{8a^2+b^2}}+b\right) \ ,\ \ 
m_{\nu 2}={\frac{1}{2}}
\left(
{\sqrt{8a^2+b^2}}-b\right) \ ,\ \ 
m_{\nu 3}=b \ .
\eqno(3.6)
$$

Since $a{\gg}b$ in the present model, we obtain
$$
\Delta m_{12}^2 = m^2_{\nu 1}-m^2_{\nu 2} =
b {\sqrt{8a^2+b^2}} \simeq 2 \sqrt{2}  ab \ ,
\eqno(3.7)
$$
$$
\Delta m_{23}^2 = m^2_{\nu 2}-m^2_{\nu 3} \simeq 2 a^2 \ ,
\eqno(3.8)
$$
so that we can predict 
$$
{\frac
{{\Delta}m_{12}^2}
{{\Delta}m_{23}^2}}
\simeq{\sqrt{2}}
{\frac{b}{a}}
=
{\sqrt{2}}
{\frac{m_e}{m_{\mu}}}=
6.7\times10^{-3} \ ,
\eqno(3.9)
$$
together with the (nearly) bimaximal mixing 
$$
U\simeq
\left(\begin{array}{ccc}
{\frac{1}{\sqrt{2}}} & -{\frac{1}{\sqrt{2}}} & 0 \\
-{\frac{1}{2}} & -{\frac{1}{2}} & {\frac{1}{\sqrt{2}}} \\
{\frac{1}{2}} & {\frac{1}{2}} & {\frac{1}{\sqrt{2}}} 
\end{array}\right).
\eqno(3.10)
$$
We regard ${\Delta}m_{12}^2$ and ${\Delta}m_{23}^2$ as ${\Delta}m_{solar}^2$ 
and ${\Delta}m_{atm}^2$, respectively.  The predicted 
value (3.9) is in excellent agreement with the observed value
\cite{Garcia,atm}
$$
\left(
{\frac
{{\Delta}m_{solar}^2}
{{\Delta}m_{atm}^2}}
\right)_{exp}\simeq
{\frac
{2.2\times10^{-5}\ {\rm eV}^2}
{3.2\times10^{-3}\ {\rm eV}^2}}
\simeq 6.9\times10^{-3}.
\eqno(3.11)
$$
Note that the neutrino mass hierarchy in the present model is $|m_{\nu 1}|
{\simeq} m_{\nu 2}{\gg} m_{\nu 3}$.  
Since ${\Delta}m_{23}^2{\simeq} m^2_{\nu 2}$, 
we can obtain the value of $m_{\nu 2}$ (and $m_{\nu 1}$) as follows
$$
|m_{\nu 1}| \simeq | m_{\nu 2}| \simeq
{\sqrt{{\Delta}m_{23}^2}}\simeq{\sqrt{{\Delta}m_{atm}^2}}\simeq5.7\times10^{-2}
\ {\rm eV}
\eqno(3.12)
$$
so that we also obtain
$$
m_{\nu 3} \simeq
{\frac{b}{{\sqrt{2}}a}} m_{\nu 2}=
{\frac{m_e}{{\sqrt{2}}m_{\mu}}} m_{\nu 2} =
2.0\times10^{-4} \ {\rm eV}.
\eqno(3.13)
$$
It will be hard to detect such small masses of $m_{\nu i}$ directly.  
Furthermore, since ${\langle}m_{\nu}\rangle\equiv|\sum_im_iU_{ei}^2|=0$ 
due to $M_{11}=0$ in the present model, 
it is also impossible to detect the effective mass 
${\langle}m_{\nu}\rangle$ in the neutrinoless double beta decay. 

{}From
$$
\Delta m^2_{atm} = \Delta m^2_{23} \simeq 2 a^2 \simeq 
\left( \frac{m_\mu}{m_\tau} f m_0 \right)^2 \ ,
\eqno(3.14)
$$
we obtain the numerical result
$$
f m_0 \simeq 2.4 \times 10^{-3} \ {\rm eV} \ ,
\eqno(3.15)
$$
$$
f \sin 2 \phi \tan\beta \ln\frac{M_1^2}{M_2^2} \simeq 1.2 \times 10^{-5} \ .
\eqno(3.16)
$$

\section{Constraints from the electroweak data}

The sensitive upper bound on $|f_{ij}|$ is , at present, given from  the 
$\mu{\rightarrow}e\overline{\nu}_e\nu_{\mu}$ decay as derived by Smirnov and 
Tanimoto \cite{Smirnov-Tanimoto}
$$
|f_{e\mu}|^2 < 2.8\times10^{-3}\ \ G_F\overline{M}^2\ ,
\eqno(4.1)
$$
where $\overline{M}$ is defined by 
$(1/\overline{M}^{2})=(1/M_1^{2})\cos^2\phi+(1/M_2^{2})\sin^2\phi$ and 
$G_F\overline{M}^2={\sqrt{2}}[\overline{M}/(v/\sqrt{2})]^2$, 
($v/\sqrt{2}=174\ {\rm GeV}$).
(Our definition of the coupling constants $f_{ij}$ are different
from that in Ref.~\cite{Smirnov-Tanimoto} by a factor 2.)
{}From the relation 
$$
\sin^2 \phi = \frac{M_\Phi^2 -M_2^2}{M_1^2 -M_2^2} \ ,
\eqno(4.2)
$$
where $M_\Phi^2$ is the squared mass of the charged scalar $\Phi^+$ 
defined by Eq.~(1.7) [the $M_{11}^2$ component of the mass matrix $M^2$ for
$(\Phi^+, h^+)$], we estimate
$$
\overline{M}^2 =\frac{M_1^2 M_2^2}{M_2^2 \cos^2\phi +M_1^2\sin^2\phi}=
\frac{M_1^2 M_2^2}{M_\Phi^2} \sim M_1^2 \ ,
\eqno(4.3)
$$
for $M_\Phi^2 \sim M_2^2 \sim 10^4$ GeV$^2$.
Therefore, we can read the constraint (4.1) as
$$
|f_{e\mu}| < \frac{M_1}{v/\sqrt{2}} \times 3\times 10^{-2} 
 \ .
\eqno(4.4)
$$
Since we expect visible effects of the Zee boson, we consider a value of 
$f_{e \mu}$ as large as possible.
For example, we suppose $f_{e \mu} \sim 10^{0}$, which means $f \sim 10^{-1}$
and $M_1 \sim 10^{4}$ GeV.
We consider that it is likely that the Zee boson has such a intermediate
mass scale.

The contribution of the Zee boson to the radiative decays
$\mu \rightarrow e \gamma$, $\tau \rightarrow e \gamma$ and
$\tau \rightarrow \mu \gamma$ are proportional to $(f_{\mu\tau}f_{e\tau})^2$, 
$(f_{\tau\mu}f_{\mu e})^2$ and $(f_{\tau e}f_{e\mu})^2$, respectively,
where
$$
(f_{\mu\tau}f_{e\tau})^2 \simeq \left(\frac{m_e}{m_\mu}\right)^2 f^4
\simeq 2.3 \times 10^{-5} f^4 \ , 
$$
$$ 
(f_{\tau\mu}f_{\mu e})^2 \simeq \left(\frac{m_e m_\mu}{m_\tau^2}\right)^2 f^4
\simeq 2.9 \times 10^{-10} f^4 \ , 
$$
$$ 
(f_{\tau e}f_{e\mu})^2 \simeq f^4 \ .
\eqno(4.5)
$$
Although the dominant mode of the radiative decay in the present model is 
$\tau\rightarrow\mu \gamma$,
the constraint on the Zee  boson contribution in the 
$\mu{\rightarrow}e\gamma$ decay is still severe compared with that in the 
$\tau{\rightarrow}\mu\gamma$, because 
the present experimental upper limits \cite{PDG00} of the partial decay widths
$\Gamma (\tau\rightarrow\mu\gamma)$ and $\Gamma (\mu\rightarrow e\gamma)$ 
are $B(\tau\rightarrow\mu\gamma)/\tau(\tau)<3.0\times10^{-6}/
2.9\times10^{-13}\ {\rm s}=1.0\times10^7\ {\rm s}^{-1}$ and 
$B(\mu{\rightarrow}e\gamma)/\tau(\mu)<4.9\times10^{-11}/2.2\times10^{-6}
\ {\rm s}=2.2\times10^{-5}\ {\rm s}^{-1}$, respectively.
However, even in the $\mu \rightarrow e \gamma$ decay, as discussed in
Refs.~\cite{Smirnov-Tanimoto} and \cite{Jarlskog}, 
the  present experimental upper limit of the decay rate 
$B(\mu{\rightarrow}e\gamma)$  cannot 
give a severe constraint on the magnitudes of $f_{ij}$.

We think that the most promising test of the present model is the
observation of a lepton-number violating decay 
$Z\rightarrow \tau^\pm \mu^\mp$, which is caused through $Z\rightarrow
\nu_e + \overline{\nu}_e$ and a triangle loop with exchange
of the Zee boson [and also through $Z\rightarrow
H_a^+ + H_a^-$ ($a=1,2$) ].
Similar lepton-number violating decays $Z\rightarrow e_i \overline{e}_j$ 
($i \neq j$) can be caused by the exchange of scalar fermions 
in a minimal SUSY standard model with explicitly broken $R$-parity via 
$L$-violation \cite{Z-e-mu}.
In the present model, it is characteristic that only the dominant
mode is $Z\rightarrow \tau^\pm \mu^\mp$,
which is proportional to
$(f_{ e\mu} f_{e\tau})^2 \simeq f^4$.
We roughly estimate 
the ratio $R=B(Z\rightarrow \tau^\pm \mu^\mp)/
B(Z\rightarrow e^+ e^-)$
as
$$
R = \frac{ \Gamma(Z\rightarrow \mu^\pm \tau^\mp)}{\Gamma(
Z\rightarrow e^+ e^-)} \sim \left( \frac{f_{e\mu} f_{e\tau}}{
16\pi^2} \ln\frac{M_1^2}{m_Z^2} \right)^2 
\sim  10^{-6} \ ,
\eqno(4.6)
$$
where we have supposed  $f_{e\mu} \sim 1$, i.e., $f\sim 10^{-1}$. 
(We must calculate all related diagrams in order to
remove the logarithmic divergence.
More details of the $Z\rightarrow e_i \overline{e}_j$ decays 
will be given elsewhere.)
The present experimental upper limit \cite{PDG00} is
$B(Z\rightarrow \tau\mu)/B(Z\rightarrow e\overline{e}) < 1.2 \times
10^{-5}/3.367\times 10^{-2}=3.6\times 10^{-4}$.
We think that the value $R \sim 10^{-6}$ is within the reach of our 
near future experiment.

Another interesting observable quantity is the mixing matrix element 
$U_{e3}$.
The direct numerical calculation from the expression (3.1) 
[not the approximate expression (3.2)] gives the value of the mixing 
matrix element 
$U_{e3}$
$$
U_{e3}=-1.64\times10^{-5}.
\eqno(4.7)
$$
However, the value (4.7) is too small to detect even in the near future,
since the present experimental upper bound \cite{CHOOZ} is
$U_{e3} < (0.22 - 0.14)$.

\section{Conclusion and discussion}

In conclusion, we have investigated a neutrino mass matrix based on the
Zee model with a badly broken horizontal symmetry.
A simple ansatz for the symmetry breaking effects leads to
$f_{e\mu}\simeq (m_\tau/m_\mu)f$, $f_{e\tau}\simeq (m_\mu/m_\tau)f$
and $f_{\mu \tau}\simeq (m_e/m_\tau)f$ for the Zee boson-lepton 
coupling constants $f_{ij}$.
The Zee mass matrix with such coupling constants $f_{ij}$ gives the
nearly bimaximal mixing (3.10) and the ratio of the squared mass 
differences $\Delta m^2_{solar}/\Delta m^2_{atm} = \sqrt{2} m_e/m_\mu 
=6.7 \times 10^{-3}$.

Since the coupling constant $c_{12}$ of the 
$\Phi_1^T i\tau_2 \Phi_2 h^-$ term in the
Lagrangian (1.1) has a dimension of mass, we think that
the Lagrangian (1.1) is not a fundamental one, but an effective one.
In the present model, the horizontal symmetry ${\rm SU(3)}_H$ is
badly broken.
We do not consider that the broken symmetry in the effective Lagrangian 
is brought by a spontaneous symmetry breaking.
We consider that there is no horizontal symmetry in the fundamental 
Lagrangian from the beginning.  Nevertheless,  
we have used the prescription of the ``broken symmetry"
only for the convenience of the phenomenological treatments.

Usually, the assumptions for the symmetry breaking are 
put on the coupling constants directly, while in the present paper,
the requirements are put on the transition matrix elements including
the coupling constants.
The present prescription is somewhat unfamiliar and strange
if quarks and leptons are fundamental entities.
The present assumption may be understood by a composite model picture
of quarks and leptons in future.  

The present phenomenological success seems to suggest that the Zee model
should be taken seriously.
Then, our future tasks will be as follows:
What is the meaning of the present prescription for the 
symmetry breaking? 
How can the Zee model we embedded into a unification scenario?

\vspace*{.2in}

\centerline{\Large\bf Acknowledgments}

We would like to thank Professor M.~Matsuda for his helpful discussions,
especially, for offering us his detailed calculation on the
$\mu \rightarrow e \gamma$ decay.
One of the authors (Y.K.) thanks  M.~Tanimoto 
for his valuable comments and  for informing useful references. 
One of the authors (A.G.) is supported by the Japan Society for Promotion
of Science (JSPS), Postdoctoral Fellowship for Foreign Researches 
in Japan  (Grant No.~99222).


\end{document}